\documentclass[a4paper,11pt]{amsart}
\usepackage{comment}

\usepackage[english]{babel}
\usepackage[T1]{fontenc}
\usepackage[utf8]{inputenc}
\usepackage{amsmath,amssymb,mathrsfs}
\usepackage{amscd}
\usepackage{tikz}
\usetikzlibrary{patterns}
\numberwithin{equation}{section}
\newtheorem{theo}{Theorem}[section]

\newtheorem{lem}{Lemma}[section]

\numberwithin{figure}{section}

\newcommand{\ch}{{\mathscr{H}}}
\newcommand{\cho}{{\mathscr{H}_0}}
\newcommand{\br}{{\mathbb{R}}}
\newcommand{\bc}{{\mathbb{C}}}
\newcommand{\spp}{{\mathcal{S}_p}}

\title[Lieb-Thirring type inequalities on complex eigenvalues]{Quantitative bounds on the discrete spectrum of non self-adjoint quantum magnetic Hamiltonians}
\author{Diomba \textsc{Sambou}}
\address{
Departamento de Matem\'aticas, Facultad de Matem\'aticas,
 Pontificia Universidad Cat\'olica de Chile, Vicu\~na Mackenna 4860, 
 Santiago de Chile}
\email{
disambou@mat.uc.cl}


\begin{document}


\begin{abstract}
We establish Lieb-Thirring type inequalities for non self-adjoint relatively compact perturbations of certain operators of mathematical physics.  We apply our results to quantum Hamiltonians of Schrödinger and Pauli with constant magnetic field of strength $b>0$. In particular, we use these bounds to obtain some information on the distribution of the eigenvalues of the perturbed operators in the neighborhood of their essential spectrum.
\end{abstract}

\maketitle







\section{Introduction and an abstract result}\label{s0}
Recently, a number of results on the spectral properties of the non self-adjoint perturbations of operators of mathematical physics were obtained. We quote the articles by Frank-Laptev-Lieb-Seiringer \cite{fra}, Borichev-Golinskii-Kupin \cite{bor}, Demuth-Katriel-Hansmann \cite{dem}, Hansmann \cite{han}, Golinskii-Kupin \cite{gol}, Pushnitskii-Raikov-Villegas-Blas \cite{pus} turned to the study of the discrete spectrum of these perturbations. The purpose of this paper is to announce and to give a brief overview of new results  in this direction. The main point is that the first part of this article describes a general construction that applies to a large class of operators containing magnetic Schrödinger, Pauli and Dirac operators of full rank with constant magnetic field, hence generalizing the methods of a recent paper by the author \cite{dio}.

Let $\cho$ be an unbounded self-adjoint operator defined on a dense subset of $L^2(\br^m), \ m\ge1$. Suppose that the spectrum $\sigma(\cho)$ of the operator is given by an infinite sequence of (real) eigenvalues of infinite multiplicity, $i.e.$
\begin{equation}\label{eq1,1}
\begin{cases}
\sigma(\cho) = \sigma_{ess}(\cho) = \bigcup_{j=0}^{\infty} \big\lbrace \Lambda_{j} \big\rbrace, \\
\textup{where $\Lambda_{0} \geq 0$, $\Lambda_{j+1} > \Lambda_{j}$, $\vert \Lambda_{j+1} - \Lambda_{j} \vert \leq \delta$, $\delta > 0$ constant.}
\end{cases}
\end{equation}
Concrete examples of operators satisfying these assumptions  are Schrödinger operators acting on $L^{2}\big(\mathbb{R}^{2d},\mathbb{C}\big)$, $d \geq 1$, and Pauli operators on $L^{2}\big(\mathbb{R}^{2d},\mathbb{C}^{2}\big)$ with constant magnetic field of strength $b > 0$, see Sections \ref{s2} and \ref{s3}, respectively. We can also consider the case of Dirac operators of full rank \big(see \cite{mel}\big). But, for simplicity, we focus on the two first examples.

On the domain of $\cho$, we consider a (non self-adjoint) relatively compact perturbation $V$ of $\cho$, and the perturbed operator
\begin{equation}\label{eq1,2}
\ch =\cho + V.
\end{equation}
This means that $\text{dom}(\cho)\subset\text{dom}(V)$, and 
$V(\cho-\lambda)^{-1}$
is compact for $\lambda \in \rho (\cho)$,  the resolvent set of the operator $\cho$.  It is well known \big(see e.g. \cite[Chapter VI]{kat}\big) that under this condition on $V$, there exists $\mu < 0$  such that 
\begin{equation}\label{eq1,3,0}
\sigma(\ch) \subset \big\lbrace \lambda \in \mathbb{C} : \hspace{0.1cm}\textup{Re} \hspace{0.3mm} \lambda \geq \mu\big\rbrace.
\end{equation}
Furthermore, we impose an additional restriction on $V$ allowing us to control the numerical range 
\begin{equation}\label{e002}
N(\ch) :=\big\lbrace \langle \ch f,f \rangle: f\in \mathrm{dom}\, (\ch), \ ||f||_{L^2(\br^m)} = 1 \big\rbrace
\end{equation}
of the operator $\ch$. Namely, 
\begin{equation}\label{e001}
\sigma(\ch)\subset \overline{N(\ch)}\subset \big\lbrace \lambda\in \bc: \mathrm{Re}\, \lambda\ge \mu_1 \big\rbrace
\end{equation}
for some $\mu_1<0$. For convenience, we put 
\begin{equation}\label{eqnu}
\mu_0=\mu_1-1.
\end{equation}

Recall that a compact operator $L$ defined on a separable Hilbert space belongs to the Schatten-von Neumann class $\spp, p\ge 1$, if
$\Vert L \Vert_{\spp} = (\textup{Tr} \hspace{0.5mm} \vert L \vert^{p})^{1/p}$ is finite. We also require that
$$
V(\cho-\lambda)^{-1}\in \spp,
$$
for some $p\ge 1$, which is a stronger condition just saying that the operator $V(\cho-\lambda)^{-1}$ is compact.

Since $V$ is a relatively compact perturbation with respect to the self-adjoint operator $\mathscr{H}_{0}$, then the Weyl criterion on the invariance of the essential spectrum implies that $\sigma_{ess}(\ch)=\sigma_{ess}(\cho) = \cup_{j=0}^{\infty} \lbrace \Lambda_{j} \rbrace$. Still, the operator $\ch$ can have a (complex) discrete spectrum $\sigma_{disc}(\ch)$ accumulating to $\cup_{j=0}^{\infty} \lbrace \Lambda_{j} \rbrace$, see Gohberg-Goldberg-Kaashoek \cite[Theorem 2.1, p. 373]{goh}, and the coming theorem gives a necessary condition on its distribution. The conclusion of the theorem is written in the form of a relation which is often called a Lieb-Thirring type inequality, see Lieb-Thirring  \cite{lt1} for original work.

\begin{theo}\label{theo1}
Let $\cho$ be a self-adjoint operator with $\sigma (\cho) = \bigcup_{j=0}^{\infty} \big\lbrace \Lambda_{j} \big\rbrace$ as above. Consider $\ch =\cho + V$, and for some $p> 1$ assume that the $V$ satisfies 
\begin{equation}\label{eq2,1}
\Vert V(\cho - \mu_{0})^{-1} \Vert_\spp^{p} \leq K_{0},
\end{equation}
with $K_{0} > 0$ constant.
Then, we have
\begin{equation}\label{esta}
\sum_{\lambda \in \sigma_{disc}(\ch)} \frac{\textup{dist} \big( \lambda,\cup_{j=0}^{\infty} \lbrace \Lambda_{j} \rbrace \big)^{p}}{\big( 1 + \vert \lambda \vert \big)^{2p}} \leq C_{0} K_{0},
\end{equation}
where $C_{0} = C(p,\mu_{0},\Lambda_{0})$ is a constant depending on $p$, $\mu_{0}$, and $\Lambda_{0}$.
\end{theo}
The proof of this theorem (see Section \ref{s11} for more details) is essentially based on a recent theorem of Hansmann \cite{han}, and a technical distortion lemma for a conformal mapping coming from complex analysis, see Lemma \ref{lem3,1}.

Applications of this result to magnetic Schrödinger operators on $L^{2} \big( \mathbb{R}^{2d},\mathbb{C} \big)$ and magnetic Pauli operators on $L^{2} \big( \mathbb{R}^{2d},\mathbb{C}^{2} \big)$ are given in Theorems \ref{theo2} and \ref{theo4}, respectively. In Golinskii-Kupin \cite{gol}, similar results are obtained for complex perturbations of finite band Schrödinger operators.

Bound \eqref{esta} can be rewritten in a simpler manner for various subsets of $\sigma_{disc}(\ch)$. For instance, let $\tau > 0$ be fixed. Then, for $\lambda$ satisfying $\vert \lambda \vert \geq \tau$, one has 
$$
\frac{1}{1 + \vert \lambda \vert} = \frac{1}{\vert \lambda \vert} \frac{1}{1 + \vert \lambda \vert^{-1}} \geq \frac{1}{\vert \lambda \vert} \frac{1}{1 + \tau^{-1}},
$$
and 
\begin{equation}\label{esta1}
\displaystyle \sum_{\substack{\lambda \hspace{0.5mm} \in \hspace{0.5mm} \sigma_{disc}(\ch) \\ \vert \lambda \vert \hspace{0.5mm} \geq \hspace{0.5mm} \tau}} \frac{\textup{dist} \big( \lambda,\cup_{j=0}^{\infty} \lbrace \Lambda_{j} \rbrace \big)^{p}}{\vert \lambda \vert^{2p}} \leq C_{1} \left( 1 + \frac{1}{\tau} \right)^{2p} K_{0}.
\end{equation}

Furthermore, if  $(\lambda_{k})\subset \sigma_{disc}(\ch)$ converges to a point of $\sigma_{ess} (\ch) = \cup_{j=0}^{\infty} \lbrace \Lambda_{j} \rbrace$, one has 
\begin{equation}\label{eq2,3}
\sum_{k} \textup{dist} \big( \lambda_{k},\cup_{j=0}^{\infty} \lbrace \Lambda_{j} \rbrace \big)^{p} < \infty.
\end{equation}
This means that, $a\, priori$, the accumulation of the eigenvalues from $\sigma_{disc}(\ch)$ in a neighborhood of a fixed $\Lambda_{j}$, $j \in \mathbb{N}$, is a monotone function of $p$.

Similarly, we can also obtain information on diverging sequences of eigenvalues $(\lambda_{k})\subset \sigma_{disc}(\ch)$. 
For example, if for some $\tau > 0$ the sequence $(\lambda_{k})$ is such that
$$
\textup{dist} \big( \lambda_{k},\cup_{j=0}^{\infty} \lbrace \Lambda_{j} \rbrace \big) \geq \tau,
$$
then one has 
\begin{equation}\label{eq2,4}
\sum_{k=1}^{\infty} \frac{1}{\vert \lambda_{k} \vert^{2p}} < \infty.
\end{equation}

We shall progress as follows. We give the sketch of the proof of our main abstract result (Theorem \ref{theo1}) in Section \ref{s11}. We apply it to magnetic $2d$-Schrödinger and $2d$-Pauli operators in Sections \ref{s2} and \ref{s3} respectively. In Section \ref{s4}, we treat the case of magnetic $(2d+1)$-Pauli operators with constant magnetic field. Here, the essential spectrum of the operator under consideration equals $\br_+$, which is rather different from the case of the essential spectrum coinciding with the (discrete) set of "Landau levels" $\cup_{j=0}^{\infty} \lbrace \Lambda_{j} \rbrace$ \eqref{eq1,1}. This requires the use of methods close to those from Sambou \cite{dio}.

We adopt mathematical physics and spectral analysis notation and terminology from Reed-Simon \cite{ree}. As for the classes of compact operators ({\it i.e.} Schatten-von Neumann ideals), we refer the reader to Simon \cite{sim} and Gohberg-Goldberg-Krupnik \cite{gohb}. Constants are generic, $i.e.$ changing from one relation to another. For a real $x$, $[x]$ denotes its integer part.

\medskip\noindent
{\it Acknowledgements.}\quad This work is partially supported by the Chilean 
Program \textit{N\'ucleo Milenio de F\'isica Matem\'atica
RC$120002$}.

\section{The abstract result: sketch of the proof}\label{s11}

The following result of Hansmann \big(see \cite[Theorem 1]{han}\big) is the first crucial point of the proof. Let $B_{0} = B_{0}^{\ast}$ be a bounded self-adjoint operator acting on a separable Hilbert space, and $B$ be a bounded operator satisfying $B - B_{0} \in  \spp$, $p > 1$. Then, we have
\begin{equation}\label{eq3,1}
\sum_{\lambda \in \sigma_{disc}(B)} \textup{dist} \big( \lambda,\sigma(B_{0})\big)^{p} \leq C \Vert B - B_{0} \Vert_\spp^{p},
\end{equation}
where the constant $C$ is explicit and depends only on $p$. Note that we cannot apply \eqref{eq3,1} to the unbounded operators $\cho$ and $\ch$. To fix this, let us consider bounded the resolvents 
\begin{equation}\label{eq3,2}
B_{0}(\mu_{0}) := (\cho - \mu_{0})^{-1} \quad \text{and} \quad B(\mu_{0}) := (\ch - \mu_{0})^{-1},
\end{equation}
where $\mu_{0}$ is the constant defined by \eqref{eqnu}. Furthermore, 
$$(\ch - \mu_{0})^{-1} - (\cho - \mu_{0})^{-1} = - (\ch - \mu_{0})^{-1} V (\cho - \mu)^{-1},
$$
and we obtain
\begin{equation}\label{eq3,3}
\big\Vert B(\mu_{0}) - B_{0}(\mu_{0}) \big\Vert_\spp^{p} \leq \big\Vert (\ch - \mu_{0})^{-1} \big\Vert^{p} \big\Vert V(\cho - \mu_{0})^{-1} \big\Vert_\spp^{p},
\end{equation}
where $|| . ||$ stays for the usual operator norm. By \eqref{e001},  we have 
$$
\sigma(\ch) \subset \overline{N(\ch)} \subset \big\lbrace \lambda \in \mathbb{C} : \hspace{0.1cm}\textup{Re} \hspace{0.3mm} \lambda \geq \mu_{1} \big\rbrace.
$$
This implies that $\textup{dist} \big(\mu_{0},\overline{N(\ch)}\big) \geq 1$, and using \cite[Lemma 9.3.14]{dav} we get
\begin{equation}\label{eq3,4}
\left\Vert (\ch - \mu_{0})^{-1} \right\Vert \leq \frac{1}{\textup{dist} \big(\mu_{0},\overline{N(\ch)}\big)} \leq 1.
\end{equation}
Consequently,  by \eqref{eq2,1}, \eqref{eq3,3} and \eqref{eq3,4}, we obtain
\begin{equation}\label{eq3,5}
\big\Vert B(\mu_{0}) - B_{0}(\mu_{0}) \big\Vert_\spp^{p} \leq K_{0},
\end{equation}
where $K_{0}$ is the constant defined in \eqref{eq2,1}. Hence, we obtain
\begin{equation}\label{eq3,6}
\sum_{z \in \sigma_{disc}(B(\mu_{0}))} \textup{dist} \big(z,\sigma(B_{0}(\mu_{0}))\big)^{p} \leq C K_{0}
\end{equation}
by applying Hansmann's theorem \eqref{eq3,1} to the resolvents $B(\mu_{0})$ and $B_{0}(\mu_{0})$. Putting $z = \varphi_{\mu_{0}}(\lambda) = (\lambda - \mu_{0})^{-1}$,
we have 
\begin{equation}\label{eq3,7}
\small{z \in \sigma_{disc} \big( B(\mu_{0}) \big) \quad \Big( z \in \sigma \big( B_{0}(\mu_{0}) \big) \Big{)} \quad \Longleftrightarrow \quad \lambda \in \sigma_{disc}(\ch) \quad \Big( \lambda \in \sigma (\cho) \Big)}.
\end{equation}
So, we come to a distortion lemma for the conformal map $z = \varphi_{\mu_{0}}(\lambda) = (\lambda - \mu_{0})^{-1}$, which is the second important ingredient of the proof of the theorem.

\begin{lem}\label{lem3,1}
Let $\mu_{0}$ be the constant defined by \eqref{eqnu}, and $\Lambda_{j}$, $j \in \mathbb{N}$, be the "Landau levels" defined by \eqref{eq1,1}. Then, the following bound holds
\begin{equation}\label{eq3,8}
\textup{dist} \big( \varphi_{\mu_{0}}(\lambda),\varphi_{\mu_{0}}(\cup_{j=0}^{\infty} \lbrace \Lambda_{j} \rbrace) \big) \geq \frac{C \hspace{0.08cm} \textup{dist} \big( \lambda,\cup_{j=0}^{\infty} \lbrace \Lambda_{j} \rbrace \big)}{\big( 1 + \vert \lambda \vert \big)^{2}}, \quad \lambda \in \mathbb{C},
\end{equation}
where $C = C(\mu_{0},\Lambda_{0})$ is a constant depending on $\mu_{0}$ and $\Lambda_{0}$. 
\end{lem}

The proof of the lemma goes as  \cite[Lemma 6.2]{dio} and is omitted. Now, combining the above lemma, estimates \eqref{eq3,6} and \eqref{eq3,7}, we get
$$
\sum_{\lambda \in \sigma_{disc}(\ch)} \frac{\textup{dist} \big( \lambda,\cup_{j=0}^{\infty} \lbrace \Lambda_{j} \rbrace \big)^{p}}{\big( 1 + \vert \lambda \vert \big)^{2p}} \leq C_{0} K_0,
$$
where $C_{0} = C(p,\mu_{0},\Lambda_{0})$ is a constant depending on $p$, $\mu_{0}$ and $\Lambda_{0}$. This concludes the proof of Theorem \ref{theo1}. \hfill $\Box$

\section{Examples: perturbations of magnetic $2d$-Schrödinger operators, $d\ge 1$}\label{s2}

Set $X_{\perp} := (x_{1},y_{1},\ldots, x_{d},y_{d}) \in \mathbb{R}^{2d}$, $d \geq 1$, and let $b > 0$ be a constant. We consider
\begin{equation}\label{eq2,6}
\displaystyle H_{0} := \sum_{j=1}^{d} \left\lbrace \left( D_{x_{j}} + \frac{1}{2}by_{j} \right)^{2} + \left( D_{y_{j}} - \frac{1}{2}bx_{j} \right)^{2} \right\rbrace,
\end{equation} 
the Schrödinger operator acting on $L^{2}\big(\mathbb{R}^{2d}\big) := L^{2}\big( \mathbb{R}^{2d},\mathbb{C} \big)$ with constant magnetic field of strength $b>0$. The self-adjoint operator $H_{0}$ is originally defined on $C_{0}^{\infty} \big( \mathbb{R}^{2d} \big)$, and then closed in $L^{2} \big( \mathbb{R}^{2d} \big)$. It is well known \big(see e.g. \cite{dim}\big) that its spectrum consists of the increasing sequence of Landau levels 
\begin{equation}\label{eq2,7}
\Lambda_{j} = b(d + 2j), \quad j \in \mathbb{N},
\end{equation}
and the multiplicity of each eigenvalue $\Lambda_{j}$ is infinite. As in \eqref{eq1,2}, define the perturbed operator
\begin{equation}\label{eq2,8}
H = H_{0} + V,
\end{equation}
where we identify the non self-adjoint perturbation $V$ with the multiplication operator by the function $V : \mathbb{R}^{2d} \rightarrow \mathbb{C}$.
Most of known results on the discrete spectrum of Schrödinger operators deal with self-adjoint perturbations $V$, and study its asymptotic behaviour at the edges of its essential spectrum. For $V$ admitting power-like or slower decay at infinity, see for instance the papers \cite[Chap. 11-12]{iv}, \cite{pus,ra,rai,sob,tam}, and for potentials $V$ decaying at infinity exponentially fast or having a compact support see \cite{raik}. For Landau Hamiltonians in exterior domains, see \cite{kac,per,push}.

We shall first consider the class of non self-adjoint electric potentials $V$ satisfying
\begin{equation}\label{eq2,9}
(\textup{Re} \hspace{0.5mm} (V)f,f) \geq \mu_{1} \Vert f \Vert^{2}
\end{equation}
for some $\mu_{1} < 0$, and the following estimate
\begin{equation}\label{eq2,10}
\vert V(X_{\perp}) \vert \leq C F(X_{\perp}), \quad F \in L^{p} \big( \mathbb{R}^{2d} \big), \quad p \geq 2,
\end{equation}
where $C > 0$ is a constant and $F$ is a positive function. The definition of the numerical range \eqref{e002} implies that
$$
\sigma(H)\subset \overline{N(H)}\subset \{\lambda\in \bc: \mathrm{Re}\, \lambda\ge \mu_1\}.
$$
Theorem \ref{theo2} is an immediate consequence of the following lemma and Theorem \ref{theo1} with $\cho = H_0$, $\ch = H$ and $m = 2d$.

\begin{lem}\label{lem4,1} \textup{\cite[Lemma 6.1]{dio}}
Let $\Lambda_{j}$, $j \in \mathbb{N}$, be the Landau levels defined by \eqref{eq2,7}, and consider $\lambda \in \mathbb{C} \setminus \cup_{j=0}^{\infty} \lbrace \Lambda_{j} \rbrace$. Assume that $F \in L^{p}\big(\mathbb{R}^{2d}\big)$, $d \geq 1$, with $p \geq 2 \big[ \frac{d}{2} \big] + 2$. Then, there exists a constant $C = C(p,b,d)$ such that
\begin{equation}\label{estk}
\left\Vert F (H_{0} - \lambda )^{-1} \right\Vert_\spp^{p} \leq \frac{C (1 + \vert \lambda \vert)^{d} \Vert F \Vert_{L^{p}}^{p}}{\textup{dist} \big( \lambda,\cup_{j=0}^{\infty} \lbrace \Lambda_{j} \rbrace \big)^{p}}.
\end{equation}
\end{lem}

\begin{theo}\label{theo2}
Let $H = H_{0} + V$ be the perturbed Schrödinger operator defined by \eqref{eq2,8} with $V$ satisfying conditions \eqref{eq2,9} and \eqref{eq2,10}. Assume that $F \in L^{p}\big(\mathbb{R}^{2d}\big)$, $d \geq 1$, with $p \geq 2 \big[ \frac{d}{2} \big] + 2$. Then, we have
\begin{equation}\label{estc}
\sum_{\lambda \in \sigma_{disc}(H)} \frac{\textup{dist} \big( \lambda,\cup_{j=0}^{\infty} \lbrace \Lambda_{j} \rbrace \big)^{p}}{\big( 1 + \vert \lambda \vert \big)^{2p}} \leq C_{1} \Vert F \Vert_{L^{p}}^{p}, 
\end{equation}
where the constant $C_{1} = C(p,\mu_{0},b,d)$ depends on $p$, $\mu_{0} := \mu_{1} - 1$, $b$ and $d$.
\end{theo}

Notice that if the electric potential $V$ is bounded, then $\mu_{0}$ can be eliminated in the constant $C_{1} = C(p,\mu_{0},b,d)$. The price we pay is the additional factor $\big( 1 + \Vert V \Vert_{\infty} \big)^{2p}$ in the RHS of \eqref{estc}, see \cite[Theorem 2.2]{dio}.

It goes without saying that we can derive relations similar to \eqref{esta1}-\eqref{eq2,4} in the present situation.

It seems appropriate to mention that the assumptions of Theorem \ref{theo2} are typically satisfied by the potentials $V : \mathbb{R}^{2d} \rightarrow \mathbb{C}$ such that
\begin{equation}\label{eq2,12}
\vert V(X_{\perp}) \vert \leq C \hspace{0.5mm} \langle X_{\perp} \rangle^{-m}, \quad m > 0, \quad p \hspace*{0.3mm} m > 2d, \quad p \geq 2,
\end{equation}
where $C > 0$ is a constant and $\langle y \rangle : = \big( 1 + \vert y \vert^{2} \big)^{1/2}$, $y \in \mathbb{R}^{n}$, $n \geq 1$. Certain examples showing the sharpness of our results in the two-dimensional case are discussed in  \cite[Subsection 2.2]{dio}.

\section{Examples: perturbations of magnetic $2d$-Pauli operators, $d\ge 1$}\label{s3}

To simplify, we consider the two-dimensional Pauli operator 
acting in $L^{2}(\mathbb{R}^{2}) := L^{2}(\mathbb{R}^{2},\mathbb{C}^{2})$ and describing a quantum non-relativistic $\frac 12$-spin-particle subject to a magnetic field of strength $b$ and electric potential $V$. The general case can be treated in a same manner (see the discussion after Theorem \ref{theo4}). The self-adjoint unperturbed Pauli operator $H_{0}$ given by 
\begin{equation}\label{eq2,14}
H_{0} := \begin{pmatrix}
   (-i\nabla - \textbf{A})^{2} - b & 0 \\
   0 & (-i\nabla - \textbf{A})^{2} + b
\end{pmatrix} =: \begin{pmatrix}
   H_{1} & 0 \\
   0 & H_{2}
\end{pmatrix},
\end{equation}
is defined originally on $C_{0}^{\infty}(\mathbb{R}^{2})$ and then closed in $L^{2}(\mathbb{R}^{2})$. Here, $\textbf{A} = (A_{1},A_{2}) : \mathbb{R}^{2} \rightarrow \mathbb{R}^{2}$ is a magnetic potential generating the magnetic field
\begin{equation}\label{eq2,15}
b(X_{\perp}) := \frac{\partial A_{2}}{\partial x} - \frac{\partial A_{1}}{\partial y}, \quad X_{\perp} = (x,y) \in \mathbb{R}^{2}.
\end{equation}
We focus on the case where $b(X_{\perp}) = b > 0$ is a constant. In this situation,  the spectrum $\sigma (H_{0})$ of the Pauli operator $H_{0}$ \big(see e.g \cite{dim}\big) is given by 
\begin{equation}\label{eq2,16}
\sigma (H_{0}) = \bigcup_{j=0}^{\infty} \big\lbrace \Lambda_{j} \big\rbrace, \quad \Lambda_{j} = 2bj.
\end{equation}

Now consider the matrix-valued electric potential 
\begin{equation}\label{eq2,17}
V(X_{\perp}) := \big\lbrace v_{\ell k}(X_{\perp}) \big\rbrace_{1 \leq \ell,k \leq 2}, \quad \quad X_{\perp} = (x,y) \in \mathbb{R}^{2},
\end{equation}
and introduce the perturbed operator 
\begin{equation}\label{eq2,18}
H = H_{0} + V,
\end{equation}
where we identify the potential $V$ with the multiplication operator by the matrix-valued function $V$. As in the case of magnetic Schrödinger operators, most of known results on the discrete spectrum of Pauli operators deal with self-adjoint perturbations $V$ \big(see e.g. \cite{raa}\big). Let us consider the class of non self-adjoint electric potentials $V$ satisfying
\begin{equation}\label{eq2,19}
(\textup{Re} \hspace{0.5mm} (V)f,f) \geq \mu_{1} \Vert f \Vert^{2},
\end{equation}
and the following estimate
\begin{equation}\label{eq2,20}
\vert v_{\ell k}(X_{\perp}) \vert \leq C F(X_{\perp}), \quad 1 \leq \ell,k \leq 2, \quad F \in L^{p} \big( \mathbb{R}^{2d} \big), \quad p \geq 2,
\end{equation}
where $C > 0$ is a constant and $F$ a positive function. Note that if the potential $V$ is diagonal, $i.e.$ $v_{12} = v_{21} = 0$, then assumption \eqref{eq2,19} is satisfied trivially if $\textup{Re} \hspace{0.5mm} (v_{11}) \geq \mu_{1}$ and $\textup{Re} \hspace{0.5mm} (v_{22}) \geq \mu_{1}$. In the case where $V$ is non-diagonal with $\textup{Re} \hspace{0.5mm} (v_{\ell k}) \geq \omega_{0}$ for some $\omega_{0} < 0$, it can be verified that assumption \eqref{eq2,19} holds with $\mu_{1} = -2|\omega_{0}|$. Furthermore, we have the following lemma giving a quantitative bound on the norm $|| V(H_0-\lambda)^{-1}||_\spp$ in terms of the $L^p$-norm of $F$. Its proof goes along the same lines as the proof of \cite[Lemma 6.1]{dio}.

\begin{lem}\label{lem5,1}
Let $d=1$, $\Lambda_{j}$, $j \in \mathbb{N}$, be the Landau levels defined by \eqref{eq2,16}, and consider $\lambda \in \mathbb{C} \setminus \cup_{j=0}^{\infty} \lbrace \Lambda_{j} \rbrace$. Assume that $F \in L^{p}\big(\mathbb{R}^{2}\big)$ with $p \geq 2$. Then, there exists a constant $C = C(p,b,d)$ such that
\begin{equation}\label{estk}
\left\Vert F (H_{0} - \lambda )^{-1} \right\Vert_\spp^{p} \leq \frac{C (1 + \vert \lambda \vert)^{d} \Vert F \Vert_{L^{p}}^{p}}{\textup{dist} \big( \lambda,\cup_{j=0}^{\infty} \lbrace \Lambda_{j} \rbrace \big)^{p}}.
\end{equation}
\end{lem}

Setting $\cho=H_0$, $\ch=H$, $m=2d=2$ and recalling Theorem \ref{theo1} readily yields the following result.

\begin{theo}\label{theo4}
Let $H = H_{0} + V$ be the perturbed Pauli operator defined by \eqref{eq2,18} with $V$ satisfying \eqref{eq2,19} and \eqref{eq2,20}. Assume that $F \in L^{p}\big(\mathbb{R}^{2}\big)$ with $p \geq 2$. Then, the following bound holds true 
\begin{equation}\label{estg}
\sum_{\lambda \in \sigma_{disc}(H)} \frac{\textup{dist} \big( \lambda,\cup_{j=0}^{\infty} \lbrace \Lambda_{j} \rbrace \big)^{p}}{\big( 1 + \vert \lambda \vert \big)^{2p}} \leq C_{3} \Vert F \Vert_{L^{p}}^{p}, 
\end{equation}
where the constant $C_{3} = C(p,\mu_{0},b,d)$ depends on $p$, $\mu_{0} := \mu_{1} - 1$, $b$ and $d$.
\end{theo}

As above, if the electric potential $V$ is bounded, then $\mu_{0}$ can be eliminated in $C_{3} = C(p,\mu_{0},b,d)$ with the additional factor $\big( 1 + \Vert V \Vert_{\infty} \big)^{2p}$ to pay in counterpart in the RHS of \eqref{estg}.

Notice that Theorem \ref{theo4} remains valid if we replace the two-dimensional Pauli operator $H_{0}$ by the general $2d$-Pauli operators acting on $L^{2}\big(\mathbb{R}^{2d},\mathbb{C}^{2}\big)$, $d \geq 1$, defined by
\begin{equation}\label{eq2,21}
H_{0} := \begin{pmatrix}
   \mathbb{H}_{0,\perp} - bd & 0 \\
   0 & \mathbb{H}_{0,\perp} + bd
\end{pmatrix} =: \begin{pmatrix}
   \mathbb{H}_{0,\perp}^{-} & 0 \\
   0 & \mathbb{H}_{0,\perp}^{+}
\end{pmatrix}.
\end{equation}
Here,
$$\displaystyle \mathbb{H}_{0,\perp} := \sum_{j=1}^{d} \left\lbrace \left( D_{x_{j}} + \frac{1}{2}by_{j} \right)^{2} + \left( D_{y_{j}} - \frac{1}{2}bx_{j} \right)^{2} \right\rbrace$$
is the $2d$-Schrödinger operator defined by \eqref{eq2,6}. In this case, the set of Landau levels is given by $\cup_{j=0}^{\infty} \lbrace \Lambda_{j} \rbrace$ with $\Lambda_j = 2bdj$, and  we require that $F \in L^{p} \big( \mathbb{R}^{2d} \big)$ with $p \geq 2 \big[ \frac{d}{2} \big] + 2$. 

Of course, the counterparts of relations \eqref{esta1}-\eqref{eq2,4} apply as well to magnetic Pauli operators under consideration.

\section{On Lieb-Thirring type inequalities for magnetic $(2d+1)$-Pauli operators, $d\ge 1$}\label{s4}

In this section, we focus on $(2d + 1)$-dimensional self-adjoint Pauli operators with constant magnetic field, acting on $L^{2} \big( \mathbb{R}^{2d+1} \big) := L^{2} \big( \mathbb{R}^{2d+1},\mathbb{C}^{2} \big)$, $d \geq 1$, defined by
\begin{equation}\label{eq1,61}
\mathbb{P}_{0} := \begin{pmatrix}
   \mathbb{H}_{0} - bd & 0 \\
   0 & \mathbb{H}_{0} + bd
\end{pmatrix} =: \begin{pmatrix}
   \mathbb{P}_{1} & 0 \\
   0 & \mathbb{P}_{2}
\end{pmatrix}.
\end{equation}
Here, as usual the constant $b > 0$ is the strength of the magnetic field. And, for the cartesian coordinates $\textup{\textbf{x}} := (x_{1},y_{1},\ldots, x_{d},y_{d},x) \in \mathbb{R}^{2d + 1}$,
$$\displaystyle \mathbb{H}_{0} := \sum_{j=1}^{d} \left\lbrace \left( D_{x_{j}} + \frac{1}{2}by_{j} \right)^{2} + \left( D_{y_{j}} - \frac{1}{2}bx_{j} \right)^{2} \right\rbrace + D_{x}^{2}, \hspace{0.5cm} D_{\nu} := -i\frac{\partial}{\partial \nu},$$
is the $(2d + 1)$-self-adjoint Schrödinger operator with constant magnetic field originally defined on $C_{0}^{\infty} \big( \mathbb{R}^{2d + 1},\mathbb{C} \big)$. It is well known \big(see e.g \cite{dim}\big) that the spectrum of the operator $\mathbb{P}_{0}$ is absolutely continuous, coincides with $[0,+\infty)$ and has an infinite set of Landau levels
\begin{equation}\label{eq1,62}
\Lambda_{j} = 2bdj, \quad j \in \mathbb{N}.
\end{equation}
We introduce the perturbed operator on the domain of the operator $\mathbb{P}_{0}$
\begin{equation}\label{eq1,63}
\mathbb{P} = \mathbb{P}_{0} + V,
\end{equation}
where we identify the perturbation $V$ with the multiplication operator by the matrix-valued function
\begin{equation}\label{eq1,64}
V(\textbf{x}) := \big\lbrace v_{\ell k}(\textbf{x}) \big\rbrace_{1 \leq \ell,k \leq 2}.
\end{equation}
We assume that $V$ is a bounded non self-adjoint perturbation such that for any $\textbf{x} \in \mathbb{R}^{2d + 1}$ and $1 \leq \ell,k \leq 2$,
\begin{equation}\label{eq1,65}
\vert v_{\ell k}(\textbf{x}) \vert \leq C F(\textup{\textbf{x}}) G(x),
\end{equation}
where $C > 0$ is a constant, $F$ and $G$ are two positive functions satisfying $F \in \big( L^{p}\cap L^{\infty} \big) (\mathbb{R}^{2d + 1})$ for $p \geq 2$, and $G \in \big( L^{2} \cap L^{\infty} \big) \big{(} \mathbb{R} \big{)}$. Under this assumption on $V$,  we obtain (see Lemma \ref{lem11}) that for any $\lambda \in \rho (\mathbb{P}_{0})$,
\begin{equation}\label{eq1,66}
\Vert F (\mathbb{P}_{0} - \lambda )^{-1} G \Vert_\spp < \infty.
\end{equation}
Once again, this implies that $V$ is a relatively compact perturbation. 

The first ingredient of the proof  is the following lemma obtained by methods similar to \cite[Lemma 3.1]{dio}.

\begin{lem}\label{lem11}
Let $d \geq 1$ and consider $\lambda \in \mathbb{C} \setminus [0,+\infty)$. Assume that $F \in \big( L^{p}\cap L^{\infty} \big) (\mathbb{R}^{2d + 1})$ with $p \geq 2 \big[ \frac{d}{2} \big] + 2$ and $G \in \big( L^{2} \cap L^{\infty} \big) \big{(} \mathbb{R} \big{)}$. Then, there exists a constant $C = C(p,b,d)$ such that 
\begin{equation}\label{est4}
\left\Vert F (\mathbb{P}_{0} - \lambda )^{-1} G \right\Vert_\spp^{p} \leq \frac{C (1 + \vert \lambda \vert)^{d + \frac{1}{2}} K_{1}}{\textup{dist} \big( \lambda,[0,+\infty) \big)^{\frac{p}{2}} \hspace{0.5mm} \textup{dist} \big( \lambda,\cup_{j=0}^{\infty} \lbrace \Lambda_{j} \rbrace \big)^{\frac{p}{4}}},
\end{equation}
where $\Lambda_{j}$, $j \in \mathbb{N}$, are the Landau levels defined by \eqref{eq1,62} and
\begin{equation}\label{eq1,50}
K_{1} := \Vert F \Vert_{L^{p}}^{p} \big{(} \Vert G \Vert_{L^{2}} + \Vert G \Vert_{L^{\infty}} \big{)}^{p}.
\end{equation} 
\end{lem}

Note that since the potential $V$ is bounded, then the numerical range of the operator $\mathbb{P}$ satisfies
\begin{equation}\label{eq1,67}
\sigma(\mathbb{P}) \subset \overline{N(\mathbb{P})} \subset \big\lbrace \lambda \in \mathbb{C} : \hspace{0.1cm}\textup{Re} \hspace{0.3mm} \lambda \geq -2 \Vert V \Vert_{\infty} \hspace{0.2cm} \textup{and} \hspace{0.2cm} \vert \textup{Im} \hspace{0.3mm} \lambda \vert \leq 2 \Vert V \Vert_{\infty}  \big\rbrace.
\end{equation}

The Lieb-Thirring type bound for the eigenvalues of the $(2d + 1)$-Pauli operator $\mathbb{P}$ is as follows.
\begin{theo}\label{theo01}
Let $\mathbb{P} = \mathbb{P}_{0} + V$ with $V$ satisfying \eqref{eq1,64} and \eqref{eq1,65}. Assume that $F \in \big( L^{p}\cap L^{\infty} \big) (\mathbb{R}^{2d + 1})$ with $p \geq 2 \big[ \frac{d}{2} \big] + 2$, $d \geq 1$, and $G \in \big( L^{2} \cap L^{\infty} \big) \big{(} \mathbb{R} \big{)}$. Define
\begin{equation}\label{eq1,005}
K := \Vert F \Vert_{L^{p}}^{p} \big{(} \Vert G \Vert_{L^{2}} + \Vert G \Vert_{L^{\infty}} \big{)}^{p} \big{(} 1 + \Vert V \Vert_{\infty} \big{)}^{d  + \frac{p}{2} + \frac{3}{2} + \varepsilon},
\end{equation} 
for $0 < \varepsilon < 1$. Then, we have
\begin{equation}\label{est0}
\displaystyle \sum_{\lambda \hspace{0.5mm} \in \hspace{0.5mm} \sigma_{disc}(H)} \frac{\textup{dist} \big( \lambda,[0,+\infty ) \big)^{\frac{p}{2} + 1 + \varepsilon} \hspace{0.5mm} \textup{dist} \big( \lambda,\cup_{j=0}^{\infty} \lbrace \Lambda_{j} \rbrace \big)^{(\frac{p}{4} - 1 + \varepsilon)_{+}}}{(1 + \vert \lambda \vert)^{\gamma}} \leq C_{5} K,
\end{equation}
where $\Lambda_{j}$, $j \in \mathbb{N}$, are the Landau levels defined by \eqref{eq1,62}, $\gamma > d + \frac{3}{2}$, and $C_{5} = C(p,b,d,\varepsilon)$ is a constant depending on $p$, $b$, $d$ and $\varepsilon$.
\end{theo}

As usual, $[x]$ denotes the integer part of $x \in \mathbb{R}$, and $x_{+} := \max (x,0)$.

\smallskip\noindent
{\it Sketch of the proof of the theorem.} The proof goes along the same lines as the proof of \cite[Theorem 2.1]{dio} with the help of Lemma \ref{lem11}. Since $\sigma_{ess} (\mathbb{P}) = [0,+\infty)$ with an infinite set of thresholds $\Lambda_{j}$, $j \in \mathbb{N}$, we obtain two types of estimates. 

First, we bound the sums depending on parts of $\sigma_{disc}(\mathbb{P})$ concentrated around a Landau level $\Lambda_{j}$ using the Schwarz-Christoffel formula \big(see e.g. \cite[Theorem 1, p. 176]{lav}\big). Namely, if we consider a rectangle
$$\Pi_{j} := \big\lbrace \lambda \in \mathbb{C} : \vert \Lambda_{j} - \hspace{0.1cm}\textup{Re} \hspace{0.3mm} \lambda \vert \leq b \hspace{0.2cm} \textup{and} \hspace{0.2cm} \vert \textup{Im} \hspace{0.3mm} \lambda \vert \leq \text{Const.} \big\rbrace$$
around a Landau level $\Lambda_{j}$, then we have
$$\small{\sum_{\lambda \in \sigma_{disc}(H) \cap \Pi_{j}} \textup{dist} \big( \lambda,[\Lambda_{0},+\infty) \big)^{\frac{p}{2} + 1 + \varepsilon} \hspace{0.5mm} \textup{dist} \big( \lambda,\cup_{j=1}^{\infty} \lbrace \Lambda_{j} \rbrace \big)^{(\frac{p}{4} - 1 + \varepsilon)_{+}} \leq C(p,b,j,\varepsilon) K},$$
with the asymptotic property $C(p,b,j,\varepsilon) \underset{j \rightarrow \infty} {\sim} j^{d+\frac{1}{2}}$. 

Second, we get the global bound summing up the previous bounds with appropriate weights as follows:
\begin{align*}
\small{\sum_{j} \hspace*{0.2cm} \frac{1}{(1 + j)^{\gamma}} \sum_{\lambda \in \sigma_{disc}(H) \cap \Pi_{j}}} & \small{\textup{dist} \big( \lambda,[\Lambda_{0},+\infty) \big)^{\frac{p}{2} + 1 + \varepsilon} \hspace{0.5mm} \textup{dist} \big( \lambda,\cup_{j=1}^{\infty} \lbrace \Lambda_{j} \rbrace \big)^{(\frac{p}{4} - 1 + \varepsilon)_{+}}} \\
& \small{\leq \hspace*{0.2cm} \sum_{j} \frac{C(p,b,d,j,\varepsilon)}{(1 + j)^{\gamma}} K}.
\end{align*}
Now, choosing $\gamma$ such that $\gamma > d + \frac{3}{2}$ and using the fact that for any $\lambda \in \Pi_{j}$ we have $1 + j \simeq 1 + \vert \lambda \vert$, we get the global bound \eqref{est0}.
\hfill $\Box$

\end{document}